\documentclass[aps,pre,floatfix,nofootinbib,showpacs]{revtex4}
\usepackage{graphicx}
\usepackage{amsmath}
\usepackage{amssymb}
\begin{document}
\author{Ramandeep S. Johal\footnote{electronic address: rsjohal@iisermohali.ac.in}}
\affiliation{Indian Institute of Science Education and Research Mohali,\\
Transit Campus: MGSIPA Complex, Sector 26, Chandigarh 160019, India}
\draft
\title{Universal efficiency at optimal work with Bayesian statistics}
\begin{abstract}
If the work per cycle of a quantum heat engine is averaged over 
an appropriate prior distribution for an external parameter $a$, the work
becomes optimal at Curzon-Ahlborn efficiency. More general priors
of the form $\Pi(a) \propto 1/a^{\gamma}$ yield optimal work at an efficiency
which stays close to CA value, in particular near equilibrium the efficiency 
scales as one-half of the Carnot value. This feature is analogous to
the one recently observed in literature for certain models of finite-time
thermodynamics. Further, the use of Bayes' theorem
implies that the work estimated with posterior probabilities also bears close
analogy with the classical formula.  These findings suggest that the notion of prior
information can be used to reveal thermodynamic features in quantum systems,
 thus pointing to a new connection  between thermodynamic behavior and the 
concept of information.
\end{abstract}
\maketitle
The connection between thermodynamics and the concept 
of information is one of the most subtle analogies in our physical theories.
It has played a central role in the exorcism of Maxwell's demon \cite{Leffbook}.  
It is also crucial to how we may understand and exploit 
quantum information \cite{Vedral2008, Ueda2009, Jacobs2008}. 
To make nanodevices \cite{Joachim2000,Serreli2007} that are functional and useful, we need to
understand their performance with regard to heat dissipation and
optimal information processing. To model such systems, standard thermodynamic processes and heat cycles
have been generalised using quantum systems as the working media \cite{Hastapoulus76}-\cite{AJM2008}.
It is well accepted that the maximal efficiency, $\eta_c = 1-T_2/T_1$, where 
$T_1 (T_2)$ is the hot (cold) bath temperature, is only obtained by a 
reversible heat engine which however involves infinitely slow processes.
For heat cycles running in a finite time, the concept
of power output becomes meaningful.  
Curzon and  Ahlborn \cite{Curzon1975}, first of all displayed an  elegant formula in the so called 
endoreversible approximation for
efficiency at maximum power, $\eta_{\rm CA} = 1-\sqrt{1-\eta_c}$.
The appearance of an optimal efficiency in different models with a value close to Curzon-Ahlborn (CA) value, 
has raised the issue of its universality
that  has  captured  the imagination  of workers in this  area since many
years  \cite{Leff}. More recently \cite{AJM2008},\cite{Broeck2005}-\cite{Segal2010},  
 a universal form for optimal efficiency, has been discussed 
within finite-time thermodynamics in the near-equilibrium regime (small value of $\eta_c$) as given by  
$\eta \approx {\eta_c}/{2} +  O( {{\eta_c}^2}) $.
 
On the other hand, in recent years, Bayesian methods of statistical inference  have 
gained popularity in physics  \cite{Dose}. 
In Bayesian probability theory, the central role is played by the concept
of prior information.  It represents state
of our knowledge about a system before any experimental data is acquired.
 The assignment of a unique distribution to a given prior information
is a non-trivial issue but may be argued on the basis of 
 maximum entropy principle and certain requirements of invariance \cite{Jaynes1968, Jaynes2003}.
Using Bayes' theorem \cite{Jeffreys1939} one can then 
 update the prior probabilities based on the new information gathered from the data.  
Recently, Bayesian methods have been applied to modular structure of networks \cite{Hofman2008},
inference of density of states \cite{Habeck2007}, 
 the interpretation of quantum probabilities \cite{Caves2002} and
other inverse problems \cite{Lemm2000}. 

In this letter, a Bayesian approach is used to show that the efficiency at optimal work for  
a quantum heat engine is related to CA value,
 after the work  per cycle is averaged over 
the prior  distribution of an external parameter. In contrast
to the finite-time models \cite{Esposito2009}, 
the heat cycle considered here is performed infinitely slowly.
The application of Bayes' theorem gives the 
optimal efficiency exactly at CA value for a whole class 
 of priors and for arbitrary  bath temperatures.
The present analysis thus provides a novel argument for the emergence of 
thermodynamic behavior in quantum heat engines from a Bayesian perspective. 
 

As a model of a heat engine, consider a quantum system with Hamiltonian $H_1 =\sum_n \varepsilon_{n}^{(1)}
\vert n\rangle \langle n\vert$, where  
energy eigenvalues  $\varepsilon_{n}^{(1)}= \varepsilon_n a_1$. The factor $\varepsilon_n$
depends on the energy level $n$ as well as other fixed parameters/constants of the
system; $a_1$ is a controllable external parameter equivalent to say, the applied magnetic field for 
a spin-1/2 system. Other examples of this class are 1-d quantum
harmonic oscillator ($a_1$ equivalent to frequency)  and a particle in 1-d box ($a_1$ inversely proportional
to the square of box-width). Initially, the quantum system is 
in thermal state $\rho(a_1) =\sum_n  p_{n}^{(1)}\vert n\rangle \langle n\vert$ at temperature 
$T_1$ with its eigenvalues given by 
the canonical probabilities $p_{n}^{(1)}$. The quantum analogue of a classical
Otto cycle between two heat baths at temperatures  $T_1$ and $T_2$ involves the following
steps \cite{Kieu2004}:
(i) the system  is detached
from the hot bath and made to undergo the  first quantum adiabatic process,
 during which the system hamiltonian changes to $H_2 =\sum_n \varepsilon_{n}^{(2)}
\vert n\rangle \langle n\vert$, where $\varepsilon_{n}^{(2)}= \varepsilon_n a_2$, 
 without any transitions between the levels and so  
the system continues to occupy its initial state. For $a_2 < a_1$, this process is the analogue of
an adiabatic expansion. The work  done {\it by} the system in this stage
is defined as the change in mean energy ${\cal W}_1=  {\rm Tr}(\rho(a_1)[H_2-H_1])$;
(ii) the system with modified energy spectrum $\varepsilon_{n}^{(2)}$ 
is brought in thermal
contact with the cold bath and it achieves a thermal state 
$\rho(a_2) =\sum_n  p_{n}^{(2)}\vert n\rangle \langle n\vert$.  The modified canonical  
probabilities $p_{n}^{(2)}$ now correspond to temperature $T_2$.
On average, heat rejected to the bath
in this stage is defined as $Q_2 = {\rm Tr}([\rho(a_2)-\rho(a_1)] H_2)$;
(iii) the system  is now detached
from the cold bath and made to undergo a second quantum adiabatic process (compression) 
during which the hamiltonian changes back to $H_1$.
Work done {\it on} the system in this stage is  ${\cal W}_2=  {\rm Tr}(\rho(a_2)[H_1-H_2])$;
(iv) finally, the system is brought in thermal contact with the hot bath again.
 Heat is absorbed by the system in this stage whence it recovers its initial 
 state and its temperature attains back the value $T_1$. 
The total work done on average in a cycle is calculated to be  
\begin{eqnarray}
{\cal W} &=& \sum_{n} \left( \varepsilon_{n}^{(1)}- \varepsilon_{n}^{(2)} \right) 
\left( p_{n}^{(1)}- p_{n}^{(2)} \right), \\
&=&  (a_1-a_2)\sum_{n} \varepsilon_{n} \left( p_{n}^{(1)}- p_{n}^{(2)} \right) >  0. 
\label{work}
\end{eqnarray}
Similarly, heat exchanged with hot bath in stage (iv) is given by
$Q_1=a_1 \sum_n \varepsilon_n  \left( p_{n}^{(1)}- p_{n}^{(2)} \right) >0.$
Heat exchanged by the system with the cold bath is
%
$Q_2 = {\cal W}-Q_1 <0$.
The efficiency of the engine  $\eta={\cal W }/Q_1$, 
is given by
\begin{equation}
\eta = 1-\left(\frac{a_2}{a_1}\right).
\label{eta}
\end{equation}
For convenience, we express ${\cal W} \equiv {\cal W}(a_1,\eta)$, using Eq. (\ref{eta}).
Consider an ensemble of such systems where now the value of 
parameter $a_1$ may vary from system to system. If the ensemble corresponds to
an actual preparation according to a certain
probability distribution $\Pi(a_1)$, then the state of the system can be expressed
as $\hat{\rho} = \int \rho(a_1)\Pi(a_1) d a_1$. Each system in the ensemble is made to perform
the quantum heat cycle described above, with a fixed efficiency $\eta$. We wish to study the optimal characteristics
of the average work, in particular  the efficiency at which
the work becomes optimal. Clearly, choice of the probability distribution $\Pi(a_1)$ is
expected to play a significant role in the conclusions. In the following, we analyse this
problem by choosing a distribution $\Pi(a_1)$  according to the prior information
available and show that the efficiency at optimal work is closely associated with CA value.


For simplicity, we now consider a two-level system
as our working medium, with $\varepsilon_0 =  0$ and   $\varepsilon_1 = 1$, so that
 the initial energy levels are $0$ and $a_1$. The work over a cycle in this case is 
\begin{equation} 
{\cal W}(a_1,\eta) = a_1 \eta  \left[ \frac{1}{\left( 1+e^{a_1/T_1}\right) } 
-  \frac{1}{\left( 1+e^{a_1(1-\eta) /T_2}\right) } \right]  > 0,
\label{worke}
\end{equation}
where Boltzmann's constant is put equal to unity. The average work with the initial state $\hat{\rho}$
for a given $\eta$, can be expressed as
\begin{equation}
\overline{W} = \int_{a_{\rm min}}^{a_{\rm max}}
{\cal  W}(a_1,\eta) \Pi (a_1) d a_1.
\label{avw}
\end{equation}
A central issue in Bayesian probability is to assign a unique prior distribution corresponding
to a given prior information.
 If the only prior information about the continuous parameter $a_1$ is that it takes positive real values
but otherwise we have complete ignorance about it,  then Jeffreys has suggested the   
prior distribution $\Pi(a_1) \propto {1}/{a_1}$ \cite{Jeffreys, Jaynes1968}, or
in a finite range, $\Pi (a_1) = \left[ {\ln \left( a_{\rm max}/ a_{\rm min}\right)} \right]^{-1}
\left(1/a_1 \right)$, where
$a_{\rm min}$ and $a_{\rm max}$ are the minimal and the maximal energy splitting achievable
for the two-level system. For the above choice, we obtain
\begin{equation} 
 \overline{W}= \left[ {\ln \left( \frac{a_{\rm max}}{ a_{\rm min}}\right)} \right]^{-1}  \eta
\left[
\frac{T_2}{(1-\eta)} \ln \left( \frac{1+ e^{a_{\rm max}(1-\eta)/T_2}} 
 {1+ e^{a_{\rm min}(1-\eta)/T_2}}  \right )
- T_1 \ln \left(\frac{1+ e^{a_{\rm max}/T_1}} {1+ e^{a_{\rm min}/T_1}}  \right )\right ].
%
\label{av1w}
\end{equation}
It can be seen  that  the average work $\overline{W}$ vanishes for
$\eta = 0$  and $\eta = \eta_c$. In between these  values of  $\eta$, the average
work exhibits a maximum.
We look for the efficiency  at which  this work becomes maximal for the given range 
$[a_{\rm min},a_{\rm max}]$, by imposing the condition  $\partial \overline{W} / \partial \eta = 0$.
Here we  consider the limit of $a_{\rm min} \to 0$ which gives 
\begin{equation} 
 \frac{T_2}{(1-\eta)^2} \ln\left[  \frac{1+ e^{a_{\rm max}(1-\eta)/T_2}}{2}\right]
-T_1 \ln\left[ \frac{1+ e^{a_{\rm max}/T_1}}{2}     \right]  -
 \frac{\eta }{(1-\eta)}  
 \frac{a_{\rm max}}{\left( 1+ e^{-a_{\rm max}(1-\eta)/T_2}\right) } =0.  
\label{etasol}
\end{equation}
The solution $\eta$ of this equation has been plotted against $a_{\rm max}$ in Fig. 1.
Interestingly, in  the asymptotic limit of $a_{\rm max} >> T_1$, the above expression reduces to  
\begin{equation}
T_1 -\frac{T_2}{(1-\eta)^2}  =0, 
\label{eqca}
\end{equation}
which yields the efficiency at optimal work as $\eta = 1- \sqrt{{T_2}/{T_1}}$, 
exactly the CA value. More significantly, the conclusion 
also holds in general i.e. for a working system with spectrum  $\varepsilon_{n}^{(1)}= \varepsilon_n a_1$ and
 with Jeffreys' prior. It is to be noted that in the asymptotic limits, the
expression for average work (Eq. (\ref{av1w})) diverges. However, 
 the limits are  taken after the derivative of work is set equal to zero in order to obtain 
 well-defined expressions for the efficiency.
\begin{figure}[ht]
\vspace{0.2cm}
\includegraphics[width=8cm]{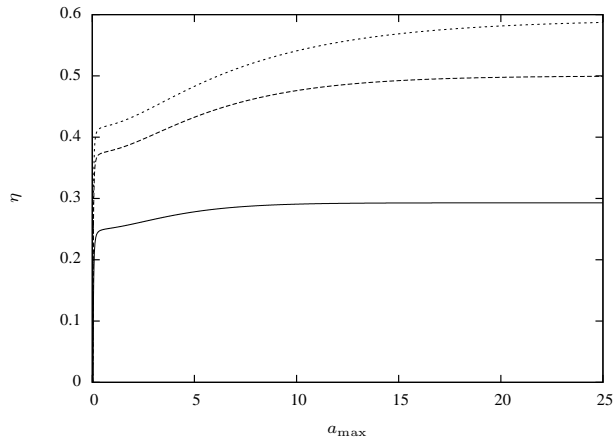}
\caption{ Efficiency versus $a_{\rm max}$ using Eq. (\ref{etasol}).
The curves correspond to $T_2 =1$ and   $T_1$  taking
values $2, 4, 6$ respectvely, from  bottom to top.
Apart from the approach to corresponding CA value at large $a_{\rm max}$, it is also
seen that the limit is approached slowly for larger temperature differences.}
\hfill
\label{fig1}
\end{figure}

It is conceivable that other choices of the prior may yield
similar results. To study consequences of deviations from the above choice,
 we consider a class of prior distributions,  $\Pi (a_1) =
{N}{{a_1}^{-\gamma}}$, defined in the range $[0,a_{\rm max} ]$,
where $N = (1-\gamma)/{a_{\rm max}}^{1-\gamma}$ and $\gamma <1$.
Upon optimisation of the average work as defined in Eq. (\ref{avw}) over $\eta$, we get
\begin{equation}
 \int_{0}^{a_{\rm max}} \left[ \frac{(a_1)^{1-\gamma}}{1+e^{a_1/T_1}} -
\frac{(a_1)^{1-\gamma}}{1+e^{a_1(1-\eta) /T_2}} \right] d a_1
-  \frac{\eta}{T_2}   \int_{0}^{a_{\rm max}} 
\frac{(a_1)^{2-\gamma}  e^{a_1(1-\eta)/T_2}  } {( 1 + e^{a_1(1-\eta) /T_2} )^2  }  d a_1 = 0.
\label{wgmodel}
\end{equation}
In the limit $a_{\rm max}$ becoming very large, the above integrals can be evaluated  using
the standard results \cite{stint}. Then the above equation is simplified to
\begin{equation}
(1-\eta^*)^{3-\gamma} -(1-\gamma)\theta^{2-\gamma}\eta^* -\theta^{2-\gamma} =0,
\label{cubeg}
\end{equation}
where $\theta = T_2/T_1$.
 Now as $\gamma \to 1$, the above equation reduces to Eq. (\ref{eqca}) and so CA value is also 
a limiting value for this model. 
Interestingly, even for other allowed values of $\gamma$, the solution $\eta^*$ of Eq. (\ref{cubeg}) 
depends only on the ratio $\theta$, apart from the parameter $\gamma$.
In particular, Laplace and Bayes have advocated a uniform prior to quantify the
state of complete ignorance.
For this case, we set $\gamma =0$. Then the above equation becomes
$(1-\eta^*)^{3} - (1 + \eta^*) \theta^{2} = 0$, which has only one real solution given as 
\begin{equation}
\eta^{*} = 1+ \frac{\theta^{4/3}}{3\left(1 +  \sqrt{1 + \frac{\theta^2}{27}}  \right)^{1/3}} - \theta^{2/3}
\left(1  + \sqrt{1 + \frac{\theta^2}{27}}  \right)^{1/3}.
\label{etacubic}
\end{equation}
This solution along with other numerical solutions of (\ref{cubeg}) for general $\gamma < 1$ are shown in Fig. 2. 
Remarkably, these curves stay very close to the CA value.
However, at this point it is not possible to say in general what prior information  may be quantified 
by the parameter $\gamma$. 
The curves in Fig. 2 are also closely similar to those observed in finite-time models at
optimal power \cite{Esposito2009}. It is seen here that in the near-equilibrium regime,
all the curves merge into each other and approach the CA value which is approximately $\eta_c/2$ in this limit. 
This can be shown as follows: taking $\theta$ to be close to unity in the near-equilibrium case,
 $\eta_c = (1-\theta)$
is close to zero. The efficiency $\eta^*$ being bounded from
above by the Carnot value is thus small too.
On using  these facts  in  the expansion  of Eq. (\ref{cubeg}), we get  
  \begin{equation}
\eta^* \approx \frac{\eta_c}{2} + \frac{(3-\gamma)}{16} {\eta_c}^2 +  O({\eta_c}^3).
\label{cubegsolution}
\end{equation}
Thus we recover the linear term ${\eta_c}/{2}$ mentioned earlier. For general values  of  $\theta$, 
the CA value is a lower bound for the efficiency at optimal work 
when $0<\gamma<1$  \cite{gammaamin}. 

\begin{figure}[ht]
\vspace{0.2cm}
\includegraphics[width=8cm]{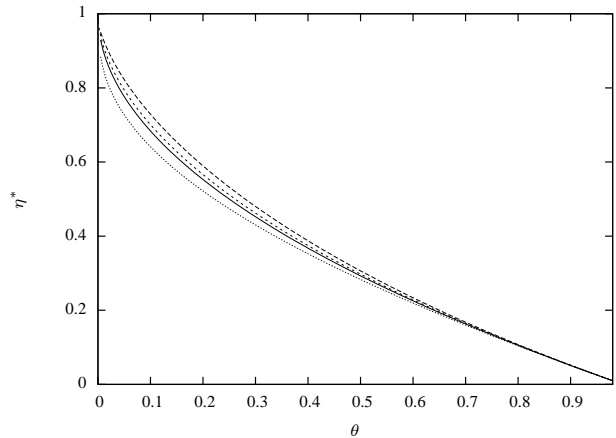}
\caption{Optimal efficiency $\eta^*$ versus $\theta = T_2/T_1$, with parameter $\gamma$ 
of the prior distribution taking
values $0, 0.25, 0.75, 1.50$ respectively, from top to bottom, excluding the solid line which 
represents the CA values, $1-\sqrt{\theta}$. For close to equilibrium ($\theta$ nearly unity),
the optimal efficiency exhibits a universal form, 
independent of $\gamma$ and given by $\eta^* \approx (1-\theta) /2 = \eta_c/2$.}
\hfill
\label{fig2}
\end{figure}
So far we have observed that 
the use of Jeffreys' prior implies that efficiency at optimal work approaches  CA value
for arbitrary bath temperatures. Further, the efficiency also approaches a universal form 
for a class of priors, for nearly equal bath temperatures. 
 In the following, we show that application of Bayes' theorem can 
restore the efficiency back to the exact CA value even for the latter choice. 
 Bayes' theorem gives a prescription to convert the prior probabilities $\Pi (a_1) d a_1$ 
into posterior probabilities. Note that during the first quantum
adiabatic process on the two-level system, the energy levels change from $(0,a_1)$ to $(0,a_2)$, but
 the system continues to occupy its initial state. 
 The respective occupation probabilities are now interpreted as conditional probabilities,
given by $p(\uparrow \vert a_1) = 1/(1+\exp(a_1/ T_1))$ and $p(\downarrow \vert a_1) = 1/(1+\exp(-a_1/ T_1))$.
If the system is found in the up ($\uparrow$) state, the work done in this step is $(a_2-a_1) = -a_1 \eta$
and the posterior probabilities are given by
\begin{equation}
p(a_1\vert \uparrow) d a_1 = \frac{p(\uparrow \vert a_1 ) \Pi (a_1) d a_1 }{\int 
p(\uparrow \vert a_1 ) \Pi (a_1) d a_1}.
\label{Btheorem}
\end{equation}
The average work for this process is now given by 
$W_1 = \int (-a_1 \eta) p(a_1\vert \uparrow) d a_1$. 
On the other hand,  if the system is found in the down ($\downarrow$)  state, the work is zero.
 Similarly, for the second quantum
adiabatic process, the work performed can be either $(+a_1 \eta)$ or 0 and the
avarage work $W_2$ for that process can be similarly calculated using the respective posterior
probabilities. Now choosing the prior  $\Pi (a_1) =  
N {{a_1}^{-\gamma}}$ and with the system being in up state, the average work for the total cycle
 $(W_1 + W_2)$, in the limit of large $a_{\rm max}$ is given by
\begin{equation}
W(\eta) = \frac{(2^{\gamma -1}-1) (1-\gamma) \zeta[2-\gamma] }
{(2^{\gamma}-1)  \zeta[1-\gamma] } \eta \left(T_1 - \frac{T_2}{(1-\eta)}  \right), 
\label{finalw}
\end{equation}
where $\gamma <1$. So using posterior probabilities, a well defined expression for average work is obtained
even if the prior is non-normalisable in the asymptotic limit.
More generally, given the value of external parameter $a_1$ and assuming canonical probabilities 
$p(n \vert a_1)$ to find the system in $n$th state, we infer the probability $p(a_1 \vert n) da_1$
about the value of $a_1$, if the system is actually found in the $n$th state.
Remarkably, the work given by eq. (\ref{finalw}) attains optimal value exactly at the CA 
efficiency, regardless of the value of $\gamma$ in the prior. 
Furthermore, the average work $W(\eta)$ 
shows the same dependence on efficiency $\eta$ as found for the classical Otto cycle in \cite{Leff}. 

In conclusion, we have argued the emergence of CA value as the efficiency at optimal work in quantum
heat engines within a Bayesian framework.
This effect of incorporating Bayesian probabilities leading to classical thermodynamic behavior 
in quantum systems has not been addressed before and may shed new light on the connection
between information and thermodynamics. Due to current interest in 
 small scale engines, the observation of similar curves (Fig. 2) as obtained in some 
recently proposed models of these engines, points to an interesting link between
finite time models and our model based on the idea of prior information.
Addressing these issues would hopefully lead to a broader perspective on
the performance characteristics of small engines and also help to understand
the limits of their performance based on principles of information. 
\section*{ACKNOWLEDEMENTS}
The author expresses his sincere thanks to Arvind, Pranaw Rungta and Lingaraj Sahu for 
interest in the work and useful discussions.


\begin{references}
\bibitem{stint} $\int_{0}^{\infty}  {(x)^{1-\gamma}}/{\left(1+e^{x/T}\right)} d x =
(1-2^{\gamma -1})  T^{2-\gamma}  \Gamma[2-\gamma] \zeta[2-\gamma]$, for
$\gamma < 2$ and where  $\Gamma[\cdot]$ and  $\zeta[\cdot]$ are the Gamma  function
and Riemann zeta functions, respectively.
\bibitem{gammaamin} If we start with the prior distribution 
$N/{a_1}^{\gamma}$  defined 
in the range $[a_{\rm min},  \infty]$, normalisable
for $\gamma  > 1$, and solve in the asymptotic limit $a_{\rm min} \to 0$, 
then CA value is an upper bound for optimal efficiency.
This choice is also presented in Fig. 2 for $\gamma = 1.5$.
\end{references}
\end{document}